M. STACHERA[1], D. STRÓZIK-KOTLORZ[2], K. SZNAJDER[1]

# MAGNETIC RESONANCE IN HUMAN BRAIN EXAMINATIONS.
## A BRIEF OUTLINE OF THE TECHNIQUES

ABSTRACT: A brief description of the magnetic resonance imaging and related advanced techniques like diffusion, perfusion and spectroscopy in the human brain examinations is given.

ABBREVIATIONS: ADC, apparent diffusion coefficient; Ala, alanine; Cho, choline; CNS, central nervous system; Cr, creatine; CSF, cerebrospinal fluid; CSI, chemical shift imaging; DW MRI, Diffusion-weighted magnetic resonance imaging; GBM, glioblastoma multiforme; GM, grey matter; HGG, high-grade gliomas; Ins, myo-inositol; Lac, lactate; LGG, low-grade gliomas; Lip, lipids; MRI, magnetic resonance imaging; MRS, magnetic resonance spectroscopy; NAA, N- acetylaspartate; NMR, nuclear magnetic resonance; PWI, perfusion imaging; SVS, single voxel spectroscopy; WM, white matter.

## 1. INTRODUCTION

Magnetic resonance is an excellent physical technique for non-invasive detection and anatomical mapping of water protons (H). The physical principles of magnetic resonance have been known since the 1940s [1,2], but the first imaging of the human body with use of MRI was performed in the 1980s. Discovery of the MR phenomenon and, then, the development of the technology allowing applications of MR were awarded several Nobel prizes.
In 1946 Felix Bloch and Edward Purcell independently discovered MR phenomenon [1,2] (Nobel Prize for Physics 1952). In 1971 Raymond Damadian showed that the nuclear magnetic relaxation times of tissues and tumors differed, thus motivating scientists to consider MR for the detection of disease. In 1975 Richard Ernst proposed MRI using phase and frequency encoding, and the Fourier Transform (Nobel Prize in Chemistry 1991). First imaging of the body using Ernst's technique was demonstrated in 1980 by Edelstein and coworkers. In 1977 Raymond Damadian demonstrated MRI called field-focusing NMR. In this same year, Peter Manseld developed the echo-planar imaging (EPI) technique, which in 1987 was used to perform real-time movie imaging of a single cardiac cycle. In 1987 Charles Dumoulin was perfecting MR angiography (MRA), which allowed imaging of flowing blood without the use of contrast agents. In 1992 functional MRI (fMRI) was developed. This technique allows the mapping of the function of the various regions of the human brain. In 2003 Paul C. Lauterbur Sir Peter Manseld were awarded the Nobel Prize in Medicine for their discoveries concerning MRI.

---

1 Department of Radiology, Regional Medical Center, Opole, Poland
2 Department of Physics, Opole University of Technology, Poland

MRI allows for structural and anatomical diagnosis in neuroradiology and since 1990s, together with advanced functional magnetic resonance techniques, like diffusion, perfusion and spectroscopy, provides valuable tools in human brain examinations. In this paper we give a short description of these techniques.

## 2. MAGNETIC RESONANCE IMAGING

Currently, MRI is the most sensitive imaging test of the brain in routine clinical practice. Most common diagnoses are: cerebrovascular accidents, neoplasms, degenerative disease, demyelinating disease like multiple sclerosis, inflammatory and infectious conditions, developmental anomalies, hydrocephalus and brain abnormalities in patients with dementia. The differential diagnosis of tumoral (neoplastic) and pseudotumoral (mainly inflammatory and demyelinating) lesions represents an important step in patient assessment, influencing further management directions (for details see eg. [3]-[9]). Initially, intracranial space-occupying lesions are typically classified as extra- or intraaxial. Discrimination of extraaxial and intraaxial brain pathology is relatively easy: intraaxial means inside brain parenchyma, extraaxial lesions are of extracerebral location, arise from outside (Fig. 1). They are usually benign. The location of brain tumors affects treatment planning and predicts their prognosis. The most common intraaxial masses include primary neoplasms (high- and low-grade gliomas), secondary (metastatic) neoplasms, lymphoma, tumefactive demyelinating lesions (TDLs), abscesses, and encephalitis. Most common extra-axial masses are meningiomas, schwannomas, arachnoid cysts, epidermoids, dermoids, chordomas, eosinophilic granulomas, pituary tumors.

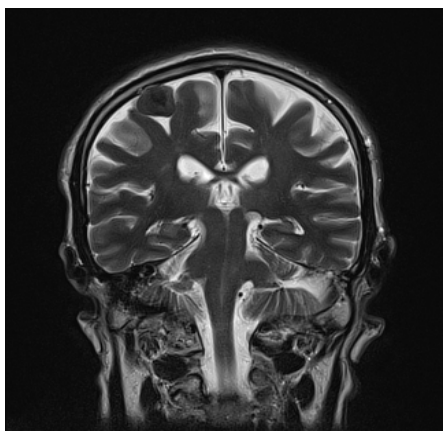

Fig. 1 Example of an extra-axial mass:
meningioma, clearly outside brain parenchyma.
Coronal T2WI.

The most common primary brain neoplasms are of glial origin, mainly astrocytomas. Low-grade (relatively benign) glial tumors (LGG) occur most often in patients aged 20–40 years. High-grade (aggresive, malignant) glial lesions (HGG) occur in older adults. Low-grade astrocytomas are mostly of pilocytic (WHO I) and fibrillary (WHO II) type. High grade astrocytoma include anaplastic (WHO III) type and malignant glioblastoma multiforme (GBM) – WHO IV. GBM is the most frequent type (50% of all astrocytomas). Another important glial tumors are oligodendrogliomas, also subdivided into low- and high grades. Meningiomas are the most common non-glial primary CNS tumors and the most common extra-axial intracranial tumors.

The MRI image interpretation begins as locating and characterizing signal abnormalities. The alterations in signal intensity depend mostly on T1, T2 and proton density of the tissues and technical parameters of pulse sequences. The higher the signal is, the brighter it will appear on the MR image. Tissue characterization includes analysis of contrast for different signal weightings (Fig.2). Water and cerebrospinal fluid with a long T1 and T2 relaxation times is dark in the T1-weighted image and bright in the T2-weighted image. The content of cystic lesions can be isointense to CSF. Brain parenchyma is of intermediate signal, on T1WI gray matter is darker than white matter (anatomical pattern), on T2WI the reverse is true. Gadolinium contrast agents reduce T1 and T2 times, resulting in an enhanced signal in the T1-weighted image. So - called conventional anatomic MRI imaging includes mostly T1-weighted, T2-weighted, FLAIR and contrast-enhanced T1-weighted images. T1 is useful primarily as depiction of anatomy, T2 is very sensitive for pathology and FLAIR is basically a T2-weighted image, but fluid spaces are attenuated and appear dark. On conventional imaging, most solid pathological lesions are hyperintensive (bright) on T2WI and hypointensive on T1WI sequences due to a high water content. Very dense and hypercellular tumors with a low water content and a high nuclear-cytoplasmasmic ratio like CNS lymphoma and PNET appear darker on T2WI. Similarily, relatively high cellularity and low water content of most meningiomas may account for the usually isointense appearance on T1W images. On T2WI they are reported to be isointense to mildly hyperintense compared with gray matter (Fig.2).

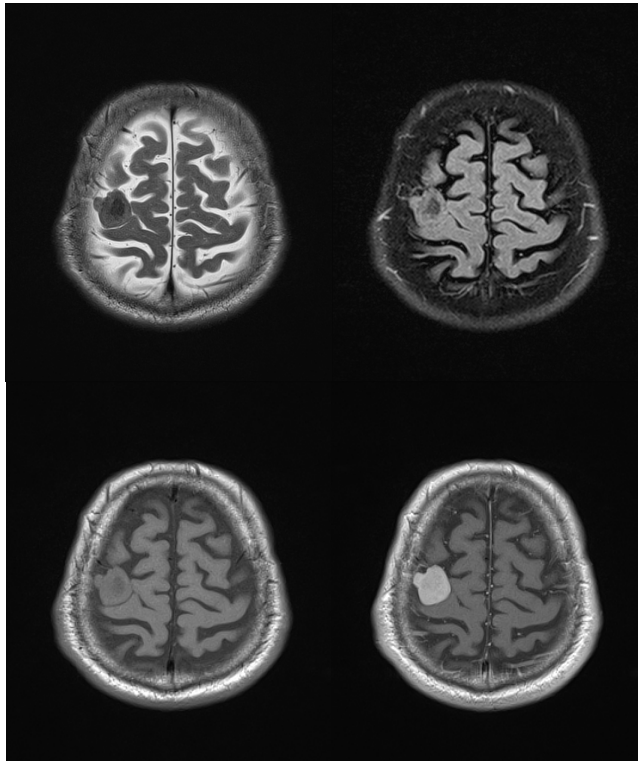

Fig. 2 Axial T2WI (upper left), FLAIR (upper right), T1WI (lower left),
contrast-enhanced T1W (lower right) of typical meningioma.

Calcifications are mostly dark on T2WI. Other causes of low intensity on T2 include blood degradation products (hemosiderin), protein, melanin, flow-void (loss of MRI signal caused by blood or CSF flow).

Most lesions have a low or intermediate signal intensity on T1WI. Exceptions to this rule are methemoglobin, high protein or lipid content of cysts, fat and melanin. Some metastases, such as melanoma, are T1 hyperintense due to the paramagnetic effects of melanin. Hemorrhagic metastases may also demonstrate T1 signal hyperintensity. Contrast enhancement represents disruption of the blood-brain barrier and is not specific. However, post-contrast T1WI improve both lesion detection (gadolinium contrast enhancement is vital to detect small metastases) and differentiation. Typically nonenhancing lesions are: low-grade neoplasms (LGG) and encephalitis. Typically enhancing lesions include abscesses, lymphomas, tumefactive demyelinating lesions (TDLs), high-grade primary neoplasms (HGG) and metastatic lesions. For peripherally enhancing masses, the main differential diagnosis lies between high-grade (HGG) and secondary brain tumours (metastases), inflammatory or demyelinating lesions (TDL) and abscesses. Nearly all meningiomas, neuromas and lymphomas enhance rapidly and intensely following contrast administration.

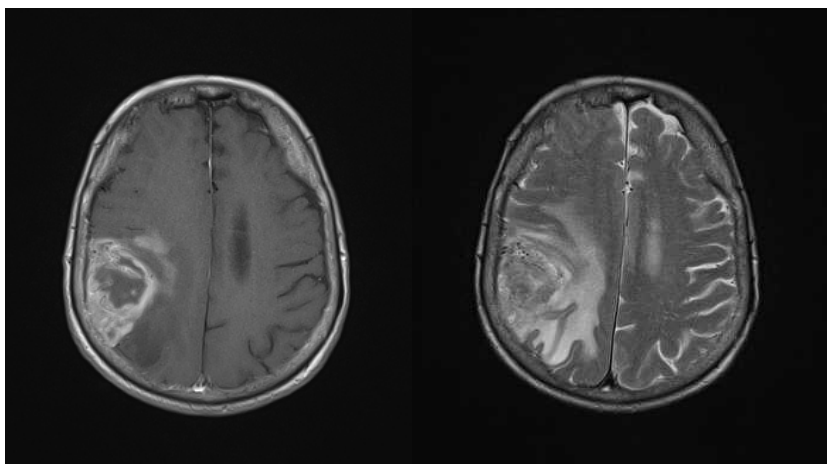

Fig.3 Glioblastoma multiforme. left: post-contrast T1WI image, right: T2WI. Central cystic area, enhancing part of the tumor and surrounding area of T2 prolongation can be appreciated.

In case of inhomogenous lesions (typical example is glioblastoma multiforme (Fig.3) it is important to analyse and describe all of the compartments: the enhancing components of the tumor, regions with peritumoral T2 prolongation and cystic areas within the tumor. The enhancing part of the tumor represents the viable tissue; cystic elements - central nonenhancing portion of ring-enhancing brain tumors represent necrosis. On gadolinium-enhanced T1WI solid part is bright and cystic is variable. The surrounding area with peritumoral T2 prolongation (bright on T2WI and FLAIR), beyond the region of contrast enhancement represents vasogenic edema, but may also contain neoplastic infiltration. In case of gliomas it is not possible to exactly delineate pure edema from tumor infiltration; precise tumor boundaries may not be evident on conventional MRI images. Metastases, on the other hand, tend not to infiltrate surrounding tissue. While interpretation of gadolinium-enhanced T1-weighted images, T2-weighted and FLAIR images remains the mainstay of brain tumor and tumor-like lesion diagnosis, this approach has limitations.

Conventional MR imaging frequently provides information that enable an accurate differential diagnosis between tumoral and pseudotumoral lesions in roughly more than 50% of cases. Imaging features include the number, topography, and morphology of lesions, intrinsic lesion architecture and the pattern of contrast medium uptake.

Integration of diagnostic information from advanced MR imaging techniques: diffusion, perfusion and spectroscopy can further improve the classification accuracy of conventional anatomic imaging.

## 3. MR DIFFUSION-WEIGHTED AND PERFUSION IMAGING

Diffusion-weighted magnetic resonance imaging (DW MRI) allows non-invasive mapping dependent on the molecular motion of water in biological tissues in vivo. As this diffusion pattern can be substantially altered by disease, DW MRI has assumed an essential role in clinical practice, especially in neuroimaging. First DW MRI images were made public in 1985 [1] and became widespread in 1990s [2].

Thermal energy possessed by water molecules is expended as continual random motion (Brownian motion), which can be quantitatively described by the self-diffusion coefficient (D). This water mobility is depend on its cellular environment (Fig. 4) and can be considered anisotropic or isotropic, as it is predominantly unidirectional or not. Low cellularity of the tissue or cell membranes disruption reflects in greater movement of water molecules, while tumors, cytotoxic edema, abscess and fibrosis can be associated with impeded diffusion.

Diffusion weighted images (DWI) are made of voxels reflect rate of water diffusion at that location. As its sensitive to intravoxel diffusion directions heterogeneity, in case of tissue dominated by isotropic water movement diffusion spectrum imaging (DSI) is useful [3]. It allows for instance mapping of axonal trajectories.

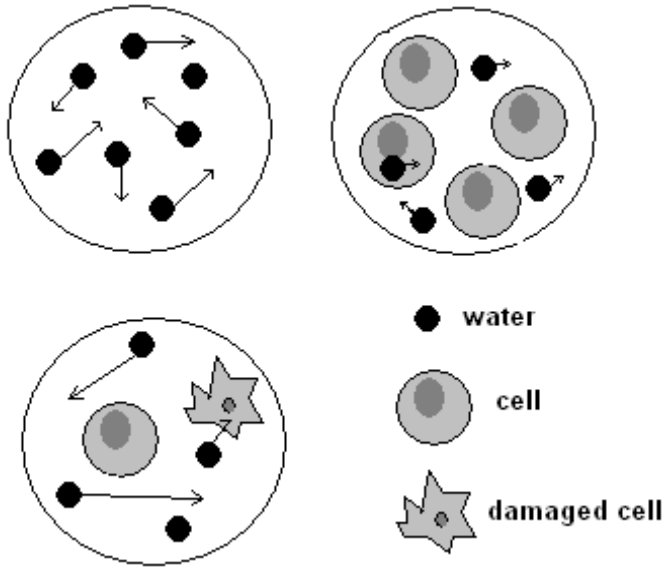

Fig. 4 Water molecule movement in different environment.

Initially DW sequence was described as adaptation of T2-weighted images [4]. Commonly used spin-echo T2-weighted sequence consist of a 90° RF pulse followed by a 180° RF pulse, with T2 decay related to transverse relaxation. In order to measure water diffusion, a dephasing gradient prior to the 180° and symmetric rephasing gradient after it must be applied. So called *b* value determines the sensitivity of DWI to process of diffusion. Higher *b* value mostly achieved by altering the gradient amplitude sequences are sensitive to even minor water displacement [5]. As DWI are also T2-weighted images, it is beneficial to separate T2 relaxation and diffusion-related changes of the signal (for instance from blood flow, perfusion, tissue pulsation). It is essential to calculate a ADC (apparent diffusion coefficient) maps. ADC is determined by the slope of the line plotted when the logarithm of relative signal intensity of tissue pointed along y-axis versus *b* value on the x-axis.

However DW MRI has been originally developed to imagine tumors of the liver, it quickly switched to stroke differential diagnosis (Fig.5). Nowadays DW MRI is commonly used for white matter disorders, especially in assessment of its tracts integrity in case of expansive lesions [6]. Within two past decades DW MRI has also been introduced for example in abdominal tumors detection and demarcation, differential diagnosis of cystic lesions and prostate cancer detecting [7].

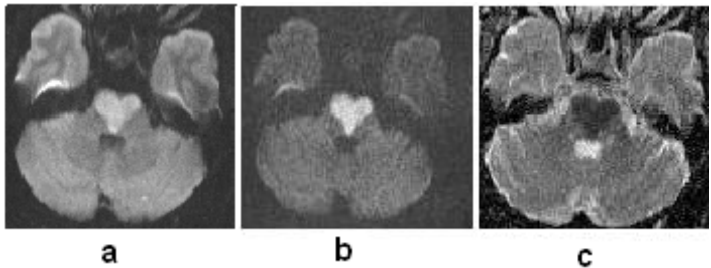

Fig. 5 Axial diffusion-weighted image at b=500 (a) and b=2000 (b)
in acute stroke of pons shows hyperintensive area in a brain stem.
This region demonstrates low signal intensity (restricted diffusion)
on ADC map (c) obtained for the same patient.

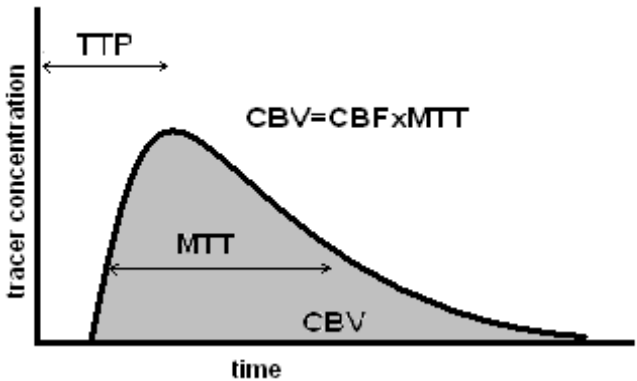

Fig. 6 Relationship of perfusion parameters demonstrated
on time-tracer concentration curve.

As a form of functional imaging MRI perfusion imaging (PWI) has become a clinical tool for evaluation of blood passage through the brain's microvascular network. Use of rapid image acquisition methods allows non-invasive assessment of various hemodynamic parameters such as cerebral blood flow (CBF), cerebral blood volume (CBV), mean transit time (MTT) and time to peak (TTP) [8] by measuring concentration of tracer agent whether exogenous (for example Gadolinium) or endogenous (spin labeling). During the first passage of Gadolinium through cerebral circulation due to drop in the signal time-to-signal curve may be processed (Fig.6). This technique exploits the magnetic susceptibility effect of the tracer with T2 or T2* weighting in echo-planar imaging sequences.

The arterial spin labeling (ASL) technique uses the saturated blood witch upstream of the slice as a tracer.

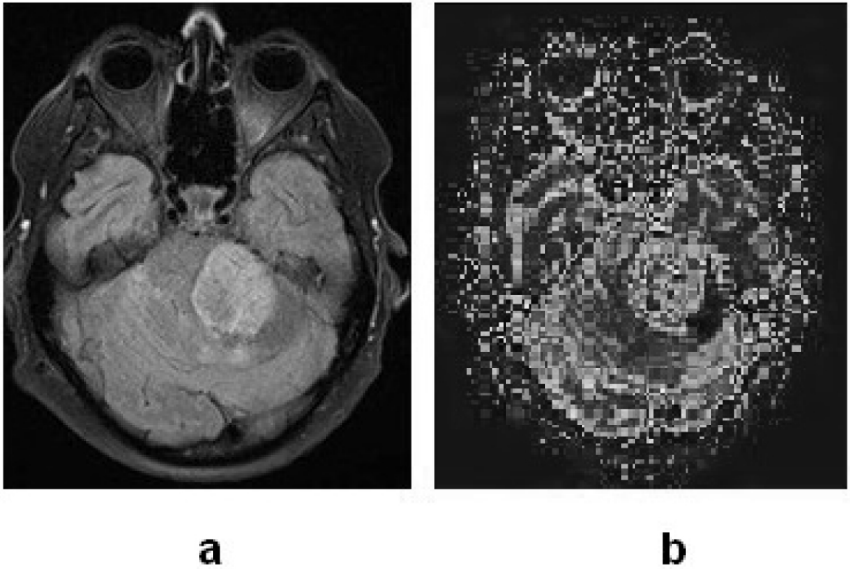

a          b

Fig. 7  56-year-old woman with pontocerebellar angle tumor.
FLAIR image shows infratentorial hyperintense mass (a),
witch reveals high cerebral blood volume (b).

The flagship application of PWI in neuroimaging has been in assessment of stroke. The ischemic brain tissue in the irreversibly damaged infarct core presents decreased CBV and CBF as well as prolonged MTT. However so called 'mismatch' between diffusion and perfusion abnormal region represents ischemic penumbra, witch is surrounding viable ischemic tissue at risk that can take benefit from thrombolytic or neuroprotective therapy [9]. Furthermore analysis of PWI and DWI parameters facilitates differential diagnosis of brain tumors like gliomas, metastases and lymphomas [10] (Fig.7). This techniques also enable to predict tumor response to antiangiogenic drugs and radiotherapy.

## 4. MAGNETIC RESONANCE SPECTROSCOPY

MRS is apart from DWI and PWI an advanced technique of the functional magnetic resonance imaging. It enables noninvasive chemical analysis of the brain, providing metabolic information. Several chemical elements can be used to obtain MRS of the brain but hydrogen is the most abundant atom in the human body. Its nucleus emits the most intense signal in an external magnetic field, in comparison to other nuclei, hence, hydrogen MRS is nowadays the most widely used method in neurospectroscopy. Together with MRI, DWI and PWI, MRS allows for the correlation of anatomical and physiological changes in the metabolic and biochemical processes occurring within given volumes in the brain.

There are two main techniques of proton MRS: single voxel and multivoxel. In SVS, a spectrum is obtained from a single sample selected volume (voxel), while in multivoxel MRS, called CSI, a much larger area can be covered, eliminating the sampling error. Multivoxel technique allow the mapping of metabolic distribution in a localized slice of the brain tissue. This can be helpful for differential diagnosis of tumors infiltrating surrounding tissue. Multivoxel spectroscopy is also used to assess response to therapy.

Basic principle of MRS is that protons in different molecules resonate at slightly different frequencies as a result of the local shielding by the electron cloud surrounding the nucleus. In this way, in a proton spectrum at the external magnetic field of 1.5 T, the metabolites are spread out between 63 and 64 MHz or between 0 and 10 ppm, respectively. Each metabolite appears at a specific ppm value, and each one reflects specific cellular and biochemical processes. Main metabolite peaks, visible in the brain spectra are NAA, Cr and Cho. NAA is a neuronal marker, Cr is a marker of energy storage and Cho is a marker of membrane turnover. For more detailed characteristic of the brain metabolites see e.g. [20-24] and ref. therein. Analysis of the obtained spectra provides biochemical information of compounds present in brain tissues, being a powerful diagnostic tool in neuroradiology [25-30]. The common way to analyze clinical spectra is to determine metabolite ratios, e.g. NAA/Cr, Cho/Cr, Cho/NAA. In comparison to the adults, newborns have much less NAA and increased Cho and Ins, which is the glial cell marker. Prominent lactate peak is visible in the newborn spectra. Progression to the adult pattern follows myelination (Fig. 8).

MRS analysis permits a safe and noninvasive examination of the brain tissue as different disease states have different MRS ''signature''. Hence MRS is useful in differential diagnosis via monitoring biochemical changes in tumors, stroke, epilepsy, metabolic disorders, infections and neurodegenenerative diseases (for review see eg. [25-30]). Specially, MRS is very sensitive for detecting brain tumors.

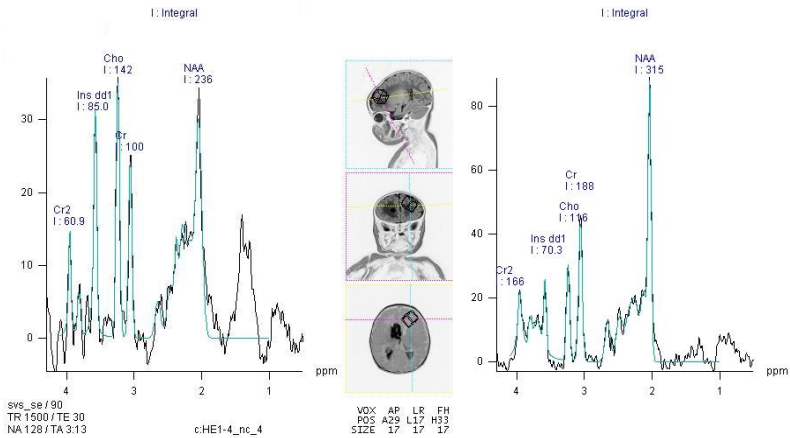

Fig. 8 Comparison of the normal MR spectrum from the frontal lobe of the newborn (left) and the adult brain (right) at TE 30 ms. Description in the text.

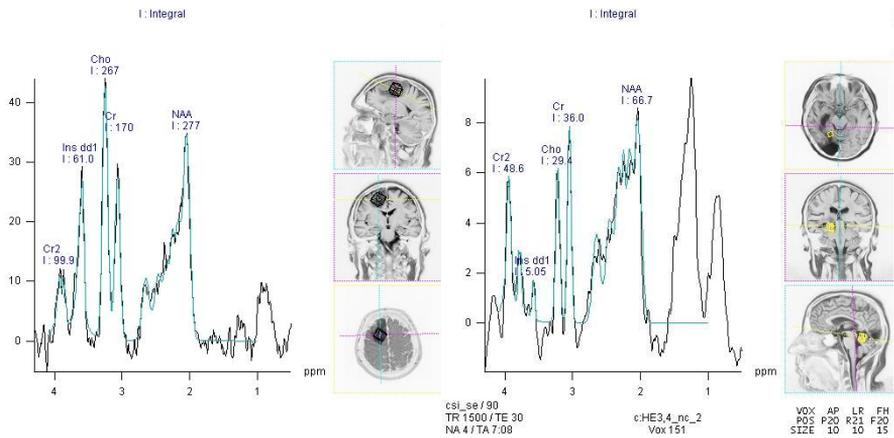

Fig. 9 Comparison of the LGG (left) and HGG GBM (right) spectra. Dominating lipid peak, denoting necrosis is visible in HGG MRS.

MRS can be helpful is determination of the malignancy grade of the brain tumor. Low grade gliomas have elevated Cho and decreased Cr and NAA levels in comparison to the healthy tissue. High grade tumors typically have more elevation of Cho and elevation of Lip and Lac as well. As malignancy increases, NAA and Cr decrease, and Cho signal increases. The detection of lipids is typical in tissue necrosis for the most common HGG, GBM. Comparison of the LGG and HGG spectra is shown in Fig. 9. MRS can also show abnormalities invisible on MRI. Namely, infiltrating gliomas, demonstrate elevated Cho beyond the region of contrast enhancement. Using this pattern, HGG can be distinguished from metastases or abscesses, which have normal Cho outside the region of enhancement. The most effective method to detect infiltration of malignant cells beyond the enhancing margins of tumor is CSI. MRS can be also helpful in diagnosis non-glial tumors, like meningioma. Meningiomas are the most common non-glial primary CNS tumors and the most common extra-axial intracranial tumors (see Section 2). The common pattern found in meningioma is slightly elevated Cho and absent or very low NAA, which denotes non neuronal origin of mass. Cr is also very low or absent and variable amounts of Lac can be found. Most important is presence of inverted Ala peak at 1.5 ppm (Fig. 10). In mostly cases, meningioma can be clearly diagnosed from the MRI characteristics, nevertheless MRS can confirm the diagnosis and moreover can be helpful in atypical cases of meningioma.

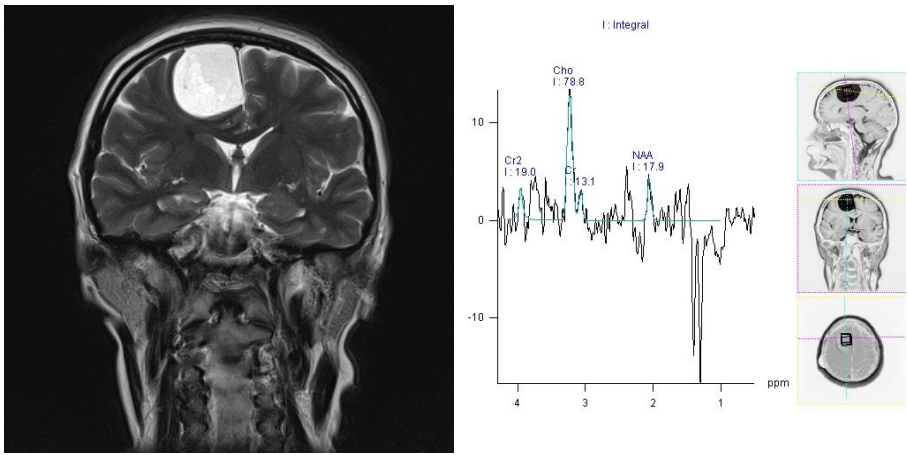

Fig. 10 Meningioma: MRI (left) and MRS (right). In MRS visible characteristic, inverted at TE 135 ms triplet Ala-Lac signal at 1.3 – 1.5 ppm .

In order to optimally distinguish between pathological and healthy tissue, one has to perform MRS examination both for tumor structure and reference normal area of the brain. Furthermore, MRS always requires correlation with the MRI and other functional imaging results before making a final reasonable neurological diagnosis.

## SUMMARY

We gave a brief characteristic of MRI and advanced neuroimaging techniques. The physical principles of magnetic resonance have been known since the 1940s but the first imaging of the human body with use of MRI was performed in the 1980s. MRI allows for structural and anatomical diagnosis in neuroradiology and since 1990s, together with advanced functional magnetic resonance techniques, like diffusion, perfusion and spectroscopy, provides valuable tools in human brain examinations. The MRI image interpretation begins as locating and characterizing signal abnormalities. The alterations in signal intensity depend mostly on T1, T2 and proton density of the tissues and technical parameters of pulse sequences. The higher the signal is, the brighter it will appear on the MR image. Tissue characterization includes analysis of contrast for different signal weightings. DW MRI allows non-invasive mapping dependent on the molecular motion of water in biological tissues in vivo. As this diffusion pattern can be substantially altered by disease, this technique is very useful in clinical practice of neuroimaging. Another form of functional imaging is PWI, which has become a clinical tool for evaluation of blood passage through the brain's microvascular network. The flagship application of PWI in neuroimaging is in assessment of stroke. MRS allows a noninvasive chemical analysis of the brain using a standard high field MR system. The major peaks in the spectrum of a normal brain include NAA, Cr, Cho and Ins, which are neuronal, energetic, membrane turnover and glial markers, respectively. In disease, two pathological metabolites can be found in the brain spectra: Lac, which is end product of anaerobic glycolysis and Lip, which is a marker of membrane breakdown, occurring in necrosis. Each disease state has its own characteristic spectroscopic image. MRS analysis always requires correlation with the MRI and other functional imaging results before making a final reasonable neurological diagnosis.

MRI and related advanced functional MR techniques, like DW MRI, PWI and MRS are nowadays a powerful tool in neuroradiology.

Acknowledgments

We thank the Technical Team of Department of Radiology at Regional Medical Center in Opole for its efforts in performing magnetic resonance examinations.